# Rigorous relations for barrier transmittance and some physical corollaries

**Sergey N. Artekha**[1, a] **and Natalya S. Artekha**[2]

[1] *Space Research Institute of RAS, Profsoyuznaya 84/32, Moscow 117997, Russia*
[2] *Peoples' Friendship University of Russia, Ulitsa Miklukho-Maklaya 6, Moscow 117198, Russia*

[a] e-mail: sergey.arteha@gmail.com  (corresponding author)

**Abstract**

Exactly solvable models are interesting for science and education, since they help in scientific search and in understanding of phenomena. Some exact solutions for simple quantum-mechanical models are considered. The models include two barriers, combinations of barrier pairs, three barriers, three wells etc. The model of two barriers can predict some interesting phenomena in the one-dimensional case. Clearing of wave and quantum-mechanical barriers (including reflection-free passage) is an important problem of physics. The rigorous equations for the transmission and reflection coefficients are derived. Barriers in substances are combined into associations, where the bond within each association is stronger than bonds between associations. Some properties of disordered media (the transparency of glasses, the conductivity of alloys or melts, the brittleness, or ductility, etc.) can be qualitatively understood from this viewpoint. The same material can exhibit various properties: transparent and opaque, metallic and non-metallic, ordered and disordered, and so on. Such transitions can occur under pressure. A model of the three-well potential can be applied to the phenomena under consideration. Some remarks on 3-D cases are made.

## 1. Introduction

The notion of a "barrier" is frequently used in various fields of science, for which a wave description can be applied, for example, in plasma physics (wave barriers), in solid-state physics (quantum-mechanical barriers), in optics, etc. The use of exactly solvable models helps in science to carry out a targeted search of new interesting phenomena, and helps to better understand the theory under study, including the student education.

From the general standpoint, two questions are of special interest for the solid state physics: 1) why does some material manifest its specific properties at standard conditions? And 2) what conditions are necessary for the material to exhibit the prescribed properties? To fully describe material properties or to solve other problems of solid state (for example, the problem of pressure influence etc.), the appropriate Schrödinger equation must be solved. This problem brings about many difficulties, and its solution was not found until now. The standard approach consists in the approximate calculation of the Schrödinger equation and in the formulation of a simple physical model [1−7].

The properties of various materials under pressure have been intensively investigated for many years till now (see classical work [8] with references herein and [9, 10]). There are many interesting phenomena in related area [11, 12]: the changes of a structure, of a color, to name but a few.

The purpose of this article is to find some rigorous expressions for the barrier transmittance proceeding from the physical meaning of appropriate physical values and to make some interesting conclusions from here. One of the purposes of this article is to draw investigators' attention to features in common in the behavior of various materials under pressure and, from simple considerations, to make some assumptions on the new effects, which can be observed. Some idea is that many materials can exhibit various properties: metallic and non-metallic, transparent and opaque, ordered and unordered, and so on.

The study of metal–insulator transitions altered the traditional separation of metals and insulators on the type of electronic spectrum and fill areas of collectivized electrons [13]. For example, for Magneli phases $V_2^{3+}V_{k-2}^{4+}O_{2k-1}$ the ions $V^{3+}$ and $V^{4+}$ are arranged randomly in the metal phase, while in the dielectric phase — orderly.

Not only the pure metal crystals (perfectly ordered, unlimited in space) can conduct electricity, but also some limited polycrystalline conductors (having obviously anisotropic structure), and even disordered alloys and liquid mercury (here not only ordering is absent, but also the fixity of the structure). Note that gaseous metals also can exhibit metallic properties [14].

We can imagine the solids as the series (regular or disordered) of barriers. The problem of the regular or disordered structure and properties of solids is of great theoretical and practical interest [13–24]. It is impossible to review all theories (the band theory of solids, the domain theory, the percolation theory, theory of topological phase transitions etc.), models (the free-electron model, the Kronig–Penney model, the cluster model, the lattice–cell model of liquid etc.) and methods (Hartree–Fock methods, the cellular method, the method LCAO, the method OPW etc.), which can be approximately applied to the area under consideration. There are many questions belonging to physics of metals, crystal physics, solid-state physics, physic of liquids and physics of disordered media, which cannot be explained in uniform manner in the context of any of these theories. Such questions are: the high conductivity of mercury and of the melts of metals, the conductivity of disordered materials (alloys, for example) outside the province of zone theory of solids, the transparency of some disordered materials (the glasses), the distinctions between the properties of various solids, and so on. The attempt to link qualitatively these various phenomena and to explain them from the unified point of view is made in this article. This is some purpose of the article. The conception of barrier associations also allows to plain and simple understand the sufficiency of some short-range order to explain a number of phenomena.

Section 2 describes exact solutions for simple 1-D models. Rigorous equations are obtained for transmission and reflection coefficients of arbitrary potential barriers in Sect. 3. Properties of a three-barrier model are considered in Sect. 4. The model of a three-well potential is described in Sect. 5. Some physical corollaries are qualitatively discussed from the simple barrier models in Sect. 6. Some features of 3-D models are discussed in Sect. 7. Section 8 contains conclusions.

**2. Exact solutions for the 1-D model of two barriers**

Although 1-D models are rather simple for real solids and they can be used with care only, these models are widely used in mesoscopic physics and physics of nanostructures, in the theory of localization etc. The basic purpose of such a consideration is to help in simple physical understanding of some effects for further finding similar effects (in more complicated form) for 2-D and 3-D real cases.

Let a wave $\psi = e^{ikx}$ impinge on some barrier. Then the reflected wave is $\psi = Re^{-ikx}$ and the transmitted wave is $\psi = Te^{ikx}$, where $T$ is the complex transmission coefficient, $R$ is the complex reflection coefficient. Let us recall that the expression for the transmittance of an individual potential barrier can be written as

$$D = |T|^2 = 1 - |R|^2. \qquad (1)$$

This expression is applicable in the absence of energy loss. Solving the Helmholtz equation for determining flows of particles (or waves), these coefficients can be found from the conditions of continuity of the wave function (in the quantum-mechanical case, for example) and its derivatives on the boundaries of the barrier.

To begin, we consider the model of two rectangular potential barriers (for some illustration only). The transmittance of the system of two identical barriers situated at a certain distance can be equal to one for a certain energy $E$. This is the first fact to which is worth paying attention and that will be used further. Resolving the Schrödinger equation (or the wave equation) for two rectangular barriers with use of the continuity conditions for the wave function and its derivative, we obtain that a resonant length $L$ for the rectangular barriers $U$ with width $a$ is determined by the formula:

$$L = \frac{2\pi n}{1{,}0798519\sqrt{E}} \pm \frac{\arccos\left\{\dfrac{U^2 - (8E^2 - 8EU + U^2)\cosh(1{,}0798519 a\sqrt{U-E})}{8E^2 - 8EU + U^2 - U^2\cosh(1{,}0798519 a\sqrt{U-E})}\right\}}{1{,}0798519\sqrt{E}}. \quad (2)$$

The energy is expressed in electron volts and the distance is expressed in angstroms in this formula only (this is made for some example only; further formulas will be expressed in the usual units). When $E > U/2$ we take the upper sign (+) and any integer number $n \geq 0$, but when $E < U/2$ we must take the lower sign (−) and any natural number $n \geq 1$ (for example, it is possible to check in a symbolic system Mathematica, that this value (2) identically makes the transmittance $D$ of the system of two barriers equals to one). We give some numerical estimation from the expression (2). Let we take $U = 0.9$ eV, $a = 2.8$ Å. For an electron with E = 0.1 eV the minimal resonant distance between barriers is equal to 14.03 Å, etc. If the electron energy is equal to 0.3 eV, the resonant distance between barriers will be $L_1 = 6.271$ Å. Next, we take, for example, $U = 1.0$ eV, $a = 2.5$ Å. When $E = 0.3$ eV the resonant distance $L_2$ between the barriers is equal to 6.462 Å. The second fact, to which is worth paying attention and that will be used further, is the following. One such "the resonant pair" could be placed at any arbitrary distance $L$ from another "a resonant pair" which is similar to the first pair, or composed of other barriers, as chosen above, but the selected energy remains resonant one for the combined disordered system. In addition, there exist other resonant energies (in addition to the chosen arbitrarily value) even when all the distances are fixed. This is illustrated in Fig. 1. It shows the total transmittance $D$ of the system consisting of two such different "resonant pairs" (the energy of 0.3 eV is randomly selected above for each pair as the resonant energy), depending on the electron energy $E$ and on the distance $L$ between these two pairs.

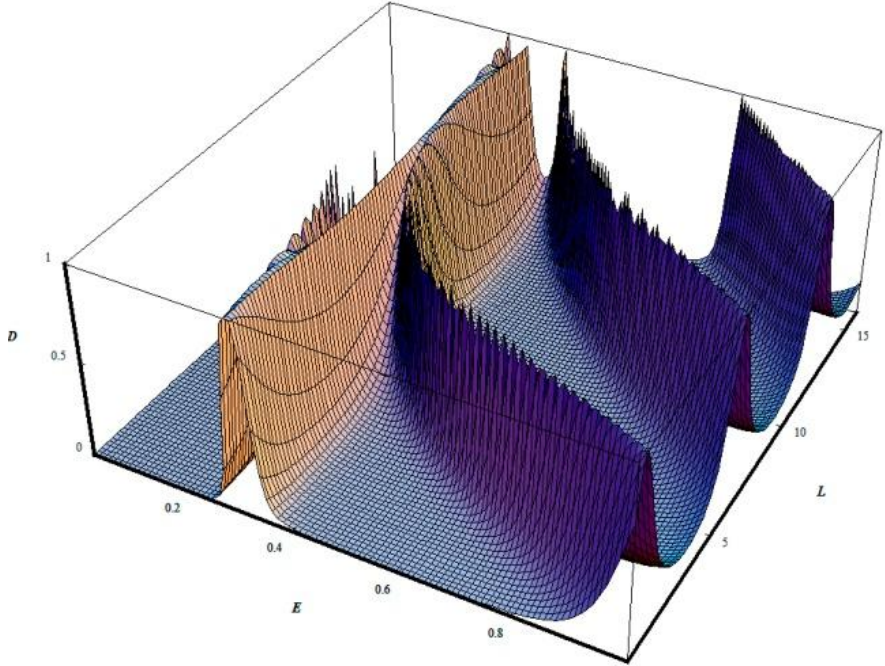

**Fig. 1.** The total transmittance of two pairs of potential barriers.

This phenomenon relates to numerous examples of so-called colored processes, when the intense, narrow separated lines are observed in the spectra of disordered media and complex chaotic processes. Some corollaries from this result will be discussed in Sect. 6.

Now let us consider the quantum-mechanical problem in the most general form. All barriers can possess any arbitrary shape. Assume that the flow of particles with mass $m$ and the energy $E$ impinges from $-\infty$ on the combined ($(R_1, T_1); (R_2, T_2)$) barrier (see Fig. 2).

Let us start with finding the complex coefficients of transmission and reflection for the combined system of barriers. In the general case the wave amplitudes for repeatedly reflected waves between the separate barriers 1 and 2 can be exactly summed up (geometric progression):

$$A = T_1\left[1 + \left(\tilde{R}_1 R_2 e^{2ikL}\right) + \left(\tilde{R}_1 R_2 e^{2ikL}\right)^2 + \cdots\right],$$

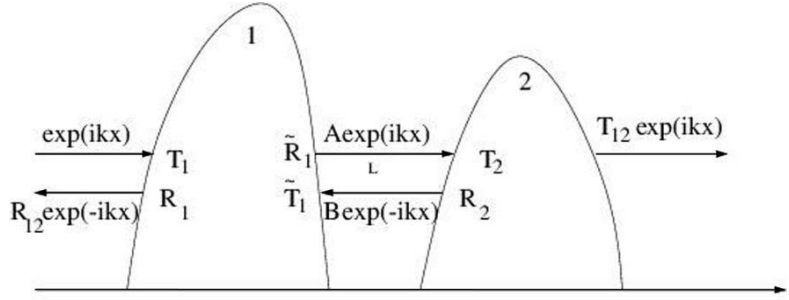

**Fig. 2.** The model of two arbitrary barriers.

and the complex transmission coefficient for the combined barrier $T_{12}$ can be found as the product $AT_2$:

$$T_{12} = \frac{T_1 T_2}{1 - \tilde{R}_1 R_2 e^{2ikL}}. \tag{3}$$

Here $\tilde{R}_1$ is the complex reflection coefficient for a wave, which impinges from $+\infty$ on barrier 1, $R_2$ is the complex reflection coefficient for a wave, which impinges from $-\infty$ on barrier 2, $T_1$ and $T_2$ are the complex transmission coefficients for waves, which impinge from $-\infty$ on barriers 1 and 2 accordingly, $L$ is the distance between the turning points for energy $E$, $k = \sqrt{2mE}/\hbar$ in the case of particles, and **k** is the wave vector in the case of waves, $\hbar$ is the Planck constant, $a_1$ and $a_2$ are the widths of barriers 1 and 2 accordingly. Similarly, summing up the return flows, we obtain the complex reflection coefficient $R_{12}$ for the combined barrier:

$$R_{12} = R_1 + \frac{T_1 R_2 \tilde{T}_1 e^{2ik(a_1+L)}}{1 - \tilde{R}_1 R_2 e^{2ikL}}. \tag{4}$$

Here we have temporarily introduced the notation $\tilde{T}_1$ of the complex transmission coefficient for a wave impinging from $+\infty$ on barrier 1. Of course, using the properties of the scattering matrix [25] due to time reversal, we could write immediately $\tilde{T}_1 = T_1$ in purely quantum mechanics (wave mechanics). However, if some generalization of quantum mechanics (wave mechanics) will be proposed by introducing irreversibility into the equations themselves, then such properties require new proof. In the most general case we must write instead of (1) the following expression:

$$|T|^2 + |R|^2 = 1 - W, \tag{1*}$$

where the value of $W$ can describe the attenuation, pumping and transformation of waves for wave mechanics, or the birth-destruction, decay, synthesis of particles in the quantum mechanical case. Obviously, we always have for the case of symmetric barriers: $T = \tilde{T}$, $R = \tilde{R}$. However, we can write in the general case only the equations $|T|^2 = |\tilde{T}|^2$, $|R|^2 = |\tilde{R}|^2$, but nothing about the values themselves. We will clarify this issue in Sect. 3.

Let us return to the usual quantum mechanics. With using symbolic mathematical programs (for example, Mathematica), expressions (3) and (4) can be strictly checked for specific barriers (identical equality for rectangular barriers or for barriers with linear grow, for example). Therefore, for the total transmittance of the combined barrier it follows, that:

$$D = |T_{12}|^2 = \frac{D_1 D_2}{\left(1 - \tilde{R}_1 R_2 e^{2ikL}\right)\left(1 - \tilde{R}_1^* R_2^* e^{-2ikL}\right)}. \tag{5}$$

We have inequality for the total transmittance:

$$\frac{1+|R_1|^2|R_2|^2-|R_1|^2-|R_2|^2}{1+|R_1|^2|R_2|^2+2|R_1\|R_2|} \leq D \leq \frac{1+|R_1|^2|R_2|^2-|R_1|^2-|R_2|^2}{1+|R_1|^2|R_2|^2-2|R_1\|R_2|},$$

and resonances can be rather sharp. They change periodically with $L$. However, of main interest here is a solution of (5) with the value $D=1$. From equality

$$D_1 D_2 = 1+|\tilde{R}_1|^2|R_2|^2 - \tilde{R}_1 R_2 e^{2ikL} - \tilde{R}_1^* R_2^* e^{-2ikL}$$

with $D_1$ and $D_2$ from (1) we obtain the following condition:

$$\tilde{R}_1 = R_2^* e^{-2ikL}. \tag{6}$$

It follows from this expression, that $D$ can be equal to 1, when $|\tilde{R}_1|^2 = |R_2|^2$. Since this condition must be carried out for some chosen value of energy $E$, it does not considerably limit the shape of barriers and the general solution. Since reflection and transmission coefficients depend continuously from $E$, always there exist possibilities of the resonant tunneling.

As a result of the periodicity of the expression (5), the change in transmittance also has a periodic character: $D(L) = D(L + \pi n/k)$, where $n = 0, 1, \ldots$. If the resonance condition (6) holds for a certain distance $L_0$, then it will be carried out for any $L_n = L_0 + \pi n/k$ also. And we have $D(L_n) = 1$. Note that until the separate barriers significantly overlap, negative integers can be also chosen as values $n$. Some corollaries from this will be considered in Sect. 6.

3. **The explicit equations for the transmission and reflection coefficients**

For definiteness, we will use the Schrödinger equation

$$\frac{\hbar^2}{2m}\frac{d^2\psi}{dx^2} + [E - U(x)]\psi = 0, \tag{7}$$

where we write $k = \sqrt{2mE}/\hbar$ for the impinge and transmitted waves ($\sim e^{ikx}$) and denote $\kappa_1(x) = \sqrt{2m(U(x)-E)}/\hbar$ (in optics: $k = \frac{\omega}{c}$, $\kappa_1(x) = \frac{\omega}{c} n(x)$, where $n(x)$ - refractive index; for plasma $n(x) = \sqrt{\varepsilon_f(x)}$, where $\varepsilon_f(x)$ - effective dielectric constant of plasma).

To derive the exact expressions, we consider the process of successive attachment of new barriers to the existed one. Let the transmission and reflection coefficients for a wave impinging from $-\infty$ on some potential barrier 1 with the width $x$ (from 0 to $x$) be $T_1(x)$ and $R_1(x)$ accordingly, and the transmission and reflection coefficients for a wave impinging from $+\infty$ on the same barrier 1 be $\tilde{T}_1(x)$ and $\tilde{R}_1(x)$ accordingly. We connect a new barrier 2 with the arbitrary height $U(x)$ and the width $\Delta x$ to the right of the barrier 1 (from $x$ to $x+\Delta x$), where $\kappa_1(x)\Delta x \ll 1$. Such a small barrier can be considered as rectangular one (and we can use the formulae from textbooks on quantum mechanics). In the first approximation, its reflection and transmission coefficients are respectively:

$$R_2(x, \Delta x) = -\frac{i(\kappa_1(x)^2 + k^2)\Delta x}{2k}, \qquad T_2(x, \Delta x) = 1 - \frac{i(\kappa_1(x)^2 + k^2)\Delta x}{2k}. \tag{8}$$

We will use the rigorous expressions (3) and (4) for the wave impinging from $-\infty$. In this case we have $L=0$, $a_2 = \Delta x$, $T_2 = \tilde{T}_2$, $R_2 = \tilde{R}_2$. We will write down in detail the analogous strong expressions for the wave impinging from $+\infty$. Then, for the combined barrier we have

$$\tilde{T}_{21}(x+\Delta x) = \frac{\tilde{T}_1(x) T_2(x, \Delta x)}{1 - \tilde{R}_1(x) R_2(x, \Delta x)}, \tag{9}$$

$$\tilde{R}_{21}(x+\Delta x) = R_2(x, \Delta x) + \frac{T_2(x, \Delta x)^2 \tilde{R}_1(x) e^{2ik\Delta x}}{1 - \tilde{R}_1(x) R_2(x, \Delta x)}. \tag{10}$$

Substituting expressions (8) into eqs. (3), (4), (9) and (10), using the definitions $\Delta T \equiv T_{12}(x+\Delta x)-T_1(x)$, $\Delta \tilde{T} \equiv \tilde{T}_{21}(x+\Delta x)-\tilde{T}_1(x)$, $\Delta R \equiv R_{12}(x+\Delta x)-R_1(x)$, $\Delta \tilde{R} \equiv \tilde{R}_{21}(x+\Delta x)-\tilde{R}_1(x)$ and passing to the limit $\Delta x \to dx \to 0$, the following explicit equations can be obtained:

$$\frac{d\ln T}{dx} = -\frac{i(\kappa_1^2+k^2)(1+\tilde{R})}{2k}, \tag{11}$$

$$\frac{dR}{dx} = -ie^{2ikx}\frac{(\kappa_1^2+k^2)T\tilde{T}}{2k}, \tag{12}$$

$$\frac{d\ln \tilde{T}}{dx} = -i\frac{(\kappa_1^2+k^2)(1+\tilde{R})}{2k}, \tag{13}$$

$$\frac{d\tilde{R}}{dx} = -i\frac{\kappa_1^2(1+\tilde{R})^2+k^2(\tilde{R}-1)^2}{2k}. \tag{14}$$

It can be checked in symbolic form on PC, that these expressions give the identities for any rectangular barriers, for any $U(x)=U_0+Cx$, etc. Comparing (13) and (11), we see, that always there is $T=\tilde{T}$, that was obvious from the properties of the scattering matrix due to time reversal. However, in the general case $R \neq \tilde{R}$, i.e. there exists the different run of phases.

If there is an empty space $L$ between barriers, then we have $\kappa_1^2 = -k^2$, therefore,

$$\frac{d\tilde{R}}{dx} = -2ik\tilde{R}, \quad i.e. \quad \tilde{R}=\tilde{R}_0 e^{-2ikL}.$$

In this case, the absolute value of the reflection coefficient of the first barrier does not change, but its phase can be made arbitrary (for a resonant connection with the second barrier).

The final equations for the complex coefficients are the following:

$$\frac{d\ln T}{dx} = -i\frac{(\kappa_1^2+k^2)(1+\tilde{R})}{2k}, \tag{15}$$

$$\frac{dR}{dx} = -ie^{2ikx}\frac{(\kappa_1^2+k^2)T^2}{2k}, \tag{16}$$

$$\frac{d\tilde{R}}{dx} + \frac{i}{2k}(\kappa_1^2+k^2)\tilde{R}^2 + \frac{i}{k}(\kappa_1^2-k^2)\tilde{R} + \frac{i}{2k}(\kappa_1^2+k^2) = 0. \tag{17}$$

In essence, the equations obtained are a certain reformulation of quantum mechanics, since they allow us to determine the wave function at any point. Equation (17) is self-sufficient, since it is not linked with other unknown quantities: it allows us to find the value of $\tilde{R}$. After that we can find the value of $T$ from (15). Also we can either immediately find the value of $R$ from (17), changing the direction of the $X$ axis to the opposite and changing the starting and ending points of the barrier, or from (16) using the found value of $T$. The equations obtained are applicable not only for continuous potentials, but also for arbitrary randomly inhomogeneous barriers (including delta-correlated processes).

If instead of eqs. (15) - (17) we want to get equations for real variables, then substituting

$$\tilde{R} = \rho e^{i\tilde{\varphi}}, \quad R = \rho e^{i\varphi}, \quad T = te^{i\delta} = \sqrt{1-\rho^2}e^{i\delta},$$

we have four unknown function of $x$: $\rho$, $\tilde{\varphi}$, $\varphi$, $\delta$. Separating the real and imaginary parts of (15)–(17), and combining these equations, we obtain the net self-consistent resulting equations and boundary conditions.

Thus, the real values $\rho$ and $\tilde{\varphi}$ can be found from the following equations:

$$\frac{d\rho}{dx} = \frac{(\kappa_1^2+k^2)}{2k}(\rho^2-1)\sin\tilde{\varphi}, \tag{18}$$

$$\rho\frac{d\tilde{\varphi}}{dx} = -\frac{(\kappa_1^2+k^2)}{2k}(\rho^2+1)\cos\tilde{\varphi} - \frac{(\kappa_1^2-k^2)}{k}\rho. \tag{19}$$

Since the value $\rho$ must start to grow at $x = 0$, the boundary conditions are the following

$$\rho(0) = 0, \quad \tilde{\varphi}(0) = -\frac{\pi}{2}. \tag{20}$$

The eqs. (18), (19) can be solved numerically in complicated cases. The first, positive point is that the results of calculations obtained for an arbitrary section are not re-counted when an additional section of the barrier is attached, but can be immediately used. Second, this is essentially some reformulation of the Schrödinger equation, since such an approach also allows us to find the value of the wave function at any arbitrary point. After obtaining these solutions, the value $\varphi$ can be found from the equation

$$\rho \frac{d\varphi}{dx} = \frac{(\kappa_1^2 + k^2)}{2k}(1-\rho^2)\cos\tilde{\varphi}, \quad \varphi(0) = -\frac{\pi}{2}. \tag{21}$$

The last variable is

$$\delta = \frac{1}{2}(\tilde{\varphi} + \varphi - 2kx + \pi), \quad \delta(0) = 0. \tag{22}$$

We also write the equation for this value:

$$\frac{d\delta}{dx} = -\frac{(\kappa_1^2 + k^2)}{2k}(1 + \rho\cos\tilde{\varphi}),$$

since we explicitly see from it that for $U(x) \geq 0$ always $d\delta/dx \leq 0$ (angle is changed clockwise). We see from (21), that the phases $\varphi$ and $\tilde{\varphi}$ begin to change synchronously: if the phase $\tilde{\varphi}$ decreases (angle is changed clockwise), then the phase $\varphi$ also decreases.

Note that if the transmitted wave $\psi \sim e^{ik_2 x}$ with $k_2 \neq k$ (the medium on both sides of the barrier is different), the transmittance of the potential barrier (by energy) can be written as:

$$D = \frac{k_2}{k}\rho^2.$$

The necessary condition of extremums for the reflection coefficient is $d\rho/dx = 0$, and, as we see from (18), it can be satisfied at $U(x_n) = 0$ (an alternation of barriers and wells) and at $\sin\tilde{\varphi}(x_m) = 0$. For the barrier system $U(x) \geq 0$, therefore, local maximums and minimums are observed at

$$\tilde{\varphi}(x_{2m-1}) = \pi + 2\pi m, \quad \tilde{\varphi}(x_{2m}) = 2\pi m, \quad m = \pm 1, \pm 2, ...$$

Similarly to the eq. (18), we can write the equation for *t*. To simplify the equations and to write them in a uniform form, we can make the substitution

$$t = \sin\alpha, \quad \rho = \cos\alpha, \quad \alpha(0) = \frac{\pi}{2}.$$

Then we have the equations:

$$\frac{d\alpha}{dx} = \frac{(\kappa_1^2 + k^2)}{2k}\sin\alpha\sin\tilde{\varphi}, \tag{18*}$$

$$\cos\alpha \frac{d\tilde{\varphi}}{dx} = -\frac{(\kappa_1^2 + k^2)}{2k}(\cos^2\alpha + 1)\cos\tilde{\varphi} - \frac{(\kappa_1^2 - k^2)}{k}\cos\alpha. \tag{19*}$$

In the case of the generalization of quantum (or wave) mechanics, instead of (8) we can write

$$R_2(x, \Delta x) = -\frac{i(\kappa_1(x)^2 + k^2)\Delta x}{2k} - w\Delta x, \quad T_2(x, \Delta x) = 1 - \frac{i(\kappa_1(x)^2 + k^2)\Delta x}{2k} - w'\Delta x. \tag{8*}$$

Then instead of (11) and (13) we get some analog of the expression (15):

$$\frac{d\ln T}{dx} = \frac{d\ln \tilde{T}}{dx} = -i\frac{(\kappa_1^2 + k^2)(1 + \tilde{R})}{2k} - (w' + w\tilde{R}). \tag{15*}$$

Thus, in the general case $T = \tilde{T}$ also, but $R \neq \tilde{R}$. Therefore, we remove the tilde-sign over *T* from further formulas.

## 4. Some remarks on three barriers

Consider the one-dimensional case of three barriers (Fig. 3). The combined barrier can be constructed from the barrier $(R_1, T_1)$ and combined barrier $(R_{23}, T_{23})$. Using formulae (3) and (4), we re-write these formulae as

$$T_{23} = \frac{T_2 T_3}{1 - \tilde{R}_2 R_3 e^{2ikL_2}}, \qquad (23)$$

$$R_{23} = R_2 + \frac{T_2^2 R_3 e^{2ik(a_2 + L_2)}}{1 - \tilde{R}_2 R_3 e^{2ikL_2}}. \qquad (24)$$

Substituting these expressions into

$$T = \frac{T_1 T_{23}}{1 - \tilde{R}_1 R_{23} e^{2ikL_1}} \qquad (25)$$

and using (1), we obtain the total transmittance:

$$D = \frac{D_1 D_2 D_3}{\left|1 - \tilde{R}_1 R_2 e^{2ikL_1} - \tilde{R}_2 R_3 e^{2ikL_2} + \tilde{R}_1 \tilde{R}_2 R_2 R_3 e^{2ik(L_1+L_2)} - \tilde{R}_1 T_2^2 R_3 e^{2ik(a_2+L_1+L_2)}\right|^2}. \qquad (26)$$

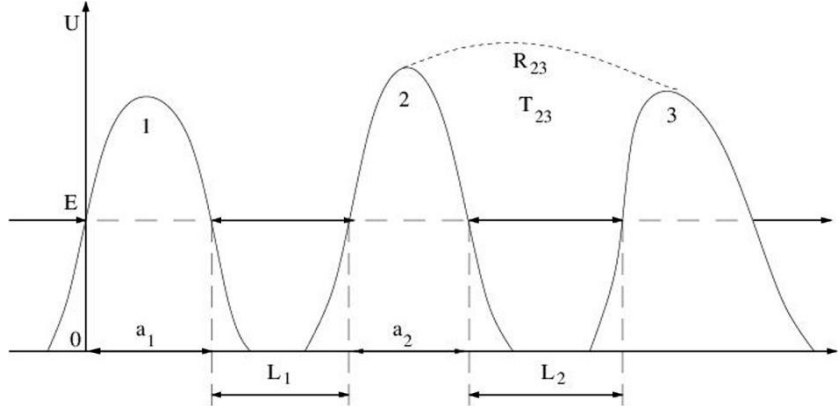

**Fig. 3.** A model of three barriers.

We see for arbitrary $L_2$ that $\min\{|R|^2\} < |R_{23}|^2 < \max\{|R|^2\}$. Therefore, if for the given energy $E$ (or for the wave vector $k$) the value of $|R_1|^2$ is situated in the same boundaries: $\min\{|R|^2\} < |R_1|^2 < \max\{|R|^2\}$, one can choose such $L_2$, that the equality $|R_1|^2 = |R_{23}|^2$ will be satisfied. In this case there exists such a distance $L_1$ that the total transmittance will be $D = 1$.

As a result of continuous character of $R$ and $T$ with $E$, the possibilities for resonant tunneling become broadened with the increasing of the number of barriers. Really, consider the combined barrier consisting of six barriers with two different forms (three equal consecutive pairs of barriers). The resonant energies will be the resonant energies for a pair of barriers plus the resonant ones for triplets. This procedure can be continued for arbitrary combined barrier consisting of any number of barriers. Increasing the number of barriers $N$, we decrease the difference between the different resonant energetic levels $\Delta E$. If this value $\Delta E$ will less than temperature fluctuations, we will observe the resonant-percolative tunneling for all structure as a whole. Therefore, not only the model of infinite periodic potential, but finite model also can describe the crystal.

We make some remarks about fluctuations. In this case, it is necessary to perform averaging in order to obtain the effective value. The question may arise that we need to average: the equation for finding the unknown value, fluctuating parameter (or additional parameters), or the value itself (after all, the rules of mathematics allow different ways to transfer a random function from one part of the equation to the other side, or from one term to another one, but the results for the effective value can be different)? We need to average measurable value, since not we average with the help of mathematical operations, but the device itself averages in the process of measuring, and it averages exactly the value that is measured. The study of convexity helps qualitatively understand the changes of the effective value due to symmetrical fluctuations (one of numerous examples is a Gaussian distribution). The study of fluctuations in the amplitude of barriers as a whole (large-scale spatial inhomogeneity) shows that the effective (average) transparency of barrier ensemble is increased for strong barriers [15]. A model of three barriers helps to determine the result of the influence of small-scale (along the spatial coordinate) fluctuations on the effective transparency. For this purpose we choose $L_1 = L_2 = 0$, $a_2 \to 0$ and consider fluctuations of this narrow central rectangular barrier in height. As a result of averaging we see that the total transparency of the combined barrier is reduced (for a more detailed explanation see [15]).

## 5. The model of three-well potential

Now we consider the own electrons with $E < 0$ (Fig. 4). If the barriers are spaced at such distance $L$, that the condition (5) is satisfied, then the electrons with energy $E$ will be collective and will produce the crystal bond. Of course, this model can be more properly referred to as "the model of three-well potential", because the energy of a binding electron is $E < 0$. Just this model of a three-well potential (not only the model of two wells [26]) can describe the change of energy levels and the change of their mutual arrangement (the phenomenon of predissociation, for example, etc.). Furthermore, we can consider the three-well potential with periodic boundary conditions instead of barrier combinations. In principle, this model can describe some phenomena with collective electrons.

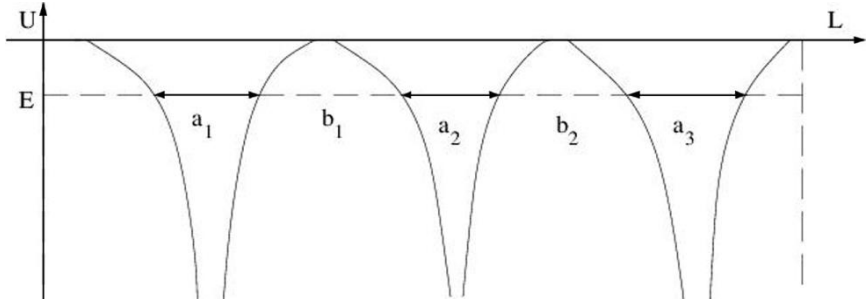

**Fig. 4.** A model of three-well potential.

Now we consider the rectangular wells for some illustration only. Let $U_i$ be the depth of the well $i$, $a_i$ denote the width of the well $i$, $b_j$ denote the width of the barrier $j$. We can find the stationary energetic levels by equating the appropriate wave functions and their derivatives on boundary of wells. Denoting

$$k = \sqrt{2mE}/\hbar, \quad k_i = \sqrt{2m(U_i - E)}/\hbar, \qquad (27)$$

and equating the determinant of the system to zero, we can obtain the condition of consistency for energy. The calculations are not complicated, but we will not write out the resulting expressions in explicit form (they can be found using symbolic computations) because of their bulkiness. We will present only plots obtained using numerical computations for specific parameters.

First, we consider some non-resonant case. Let $a_2 = a_1 = 8 \cdot 10^{-8}$ cm, $a_3 = 10^{-7}$ cm, $U_1 = 5 \cdot 10^{-12}$ erg, $U_3 = U_2 = 8 \cdot 10^{-12}$ erg, $b_2 = 3.5 \cdot 10^{-8}$ cm. All wells are different. Let now we change $b_1$ only. The dependence for energetic levels on $b_1$ at the range of $[0, U_1]$ is shown in Fig. 5 (it is typical for such cases). Here we marked the value of $b_{01} = 2.8 \cdot 10^{-8}$ cm at the plot for definiteness only.

We see that levels are negligibly shifted as long as $b_1 \geq b_{01}$, *i.e.* shifts between associations lead to relatively insignificant changes in properties of the system as a whole. Some changes in the levels are observed only when the width of the barrier separating the wells tends to zero (when the wells tend to combining). However, any consistent energy can be associated (by intersection) with no more than one energetic level (see Fig. 5). Consistent energies cover small part of the range of $[0, U_1]$.

Now we consider the resonant case, when the width of the barrier $b_2$ equals to the value $b_1$, and the further changing of these barrier widths occurs coherently ($b_1 = b_2 \equiv b$). We consider the simple case of the identical rectangular wells with $a_3 = a_2 = a_1 = 9 \cdot 10^{-8}$ cm, $U_3 = U_2 = U_1 = 5 \cdot 10^{-12}$ erg. In this case, the condition of consistency for energy leads to other energetic levels. A typical dependence for energetic levels is shown in Fig. 6.

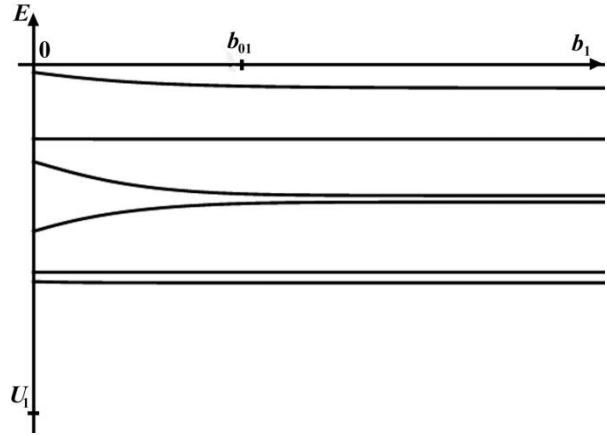

**Fig. 5.** The dependence of levels on the distance $b_1$.

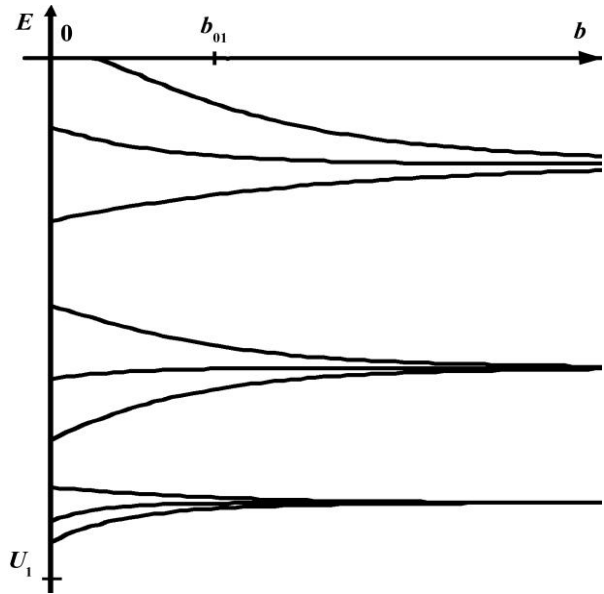

**Fig. 6.** The dependences of levels on the distance $b$.

We can see that levels are remarkable shifted with changing of $b$, *i.e.* shifts inside associations can lead to relatively significant changes in properties of the system as a whole. Levels can appear, or disappear with changing of the width of the separating barriers. Row of consistent energies can be associated (by intersection) with more than one energetic level (see Fig. 6). Consistent energies cover the most part of the range of $[0, U_1]$. Such cases can describe the transitions between materials with different properties.

## 6. Some hypotheses for the 1-D model

In this Sect., we will consider interesting qualitative corollaries from the results of previous Sects. Assuming that the distance between the barriers $L$ varies (decreases, for example), we shall determine the transmittance of a combined barrier for the flow of externally impinging particles (or waves). It is clear that the transmittance must vary periodically with *L*. Therefore, for a one-dimensional crystal, which can be described as a sequence of the same equidistant barriers (or potential wells), the quasi-periodic variations of a crystal transmittance for external particles (or waves) with *L* can be observed. The region of such quasi-periodic changes is bounded, because the separate barriers can overlap. As a result, all changes will be observed as consecutive ones (nearly periodical, for example). The transmittance will also oscillate with the energy *E* of externally impinging particles (or waves). It is evident from (5), that the period of these *L*-variations depends on the energy *E* of particles (or waves). Besides, it follows from the *E*-dependence, that the color of a crystal can be changed with *L* (under pressure, for example) and some specific hues appear and disappear consecutively (with *L*) in a crystal.

Now we consider own electrons of a crystal. In the case of metal, the collective electrons can provide the existence of a metallic bond. In addition, the collective electrons of a partially filled band provide the conductivity of this metal. The collective electrons can be associated with some resonant energies. The bands in non-metallic crystal must shift with *L*, and the energy of electrons can become resonant, that is, the nonmetal − metal transitions can occur. (We note in the brackets that even transitions to the superconducting state are possible under pressure [27]) Since new levels at some energies will shift, appear and disappear for own (internal) electrons, the dependence of conductivity will piecewise quasi-periodical with L (sometimes with breaks - some jumps are possible). Therefore, the pressure dependence of conductivity $\sigma(P)$ can be oscillating. Certainly, for more rigorous description of phase transitions in solids, we must consider the band transformations (that can be understood based on modern *ab initio* simulations). In this case, the barrier width $b_i$ will decrease under pressure (but not the well width - the distance between barriers $a_i$). For some illustration, we consider some periodic potential from the following rectangular barriers and the corresponding energetic levels - bands (see Fig. 7).

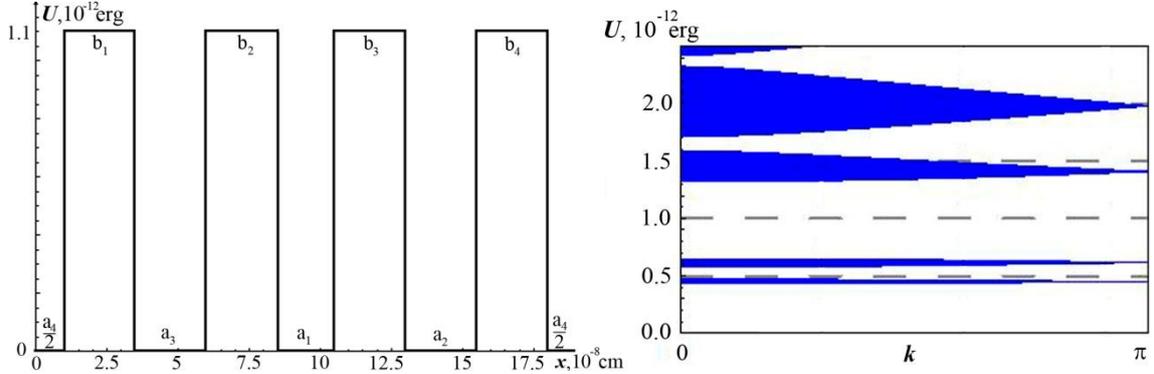

**Fig. 7.** The periodic cell and corresponding energetic bands.

Here $U_i = 1.1 \cdot 10^{-12}$ erg, $b_i = 2.5 \cdot 10^{-8}$ cm, $i = 1,2,3,4$, $a_4 = a_1 = 2 \cdot 10^{-8}$ cm, $a_3 = a_2 = 2.5 \cdot 10^{-8}$ cm.

An increase in pressure leads to a decrease in $b_i$ values. For some illustration, Fig. 8 demonstrates the energy bands at 40% compression ($b_i \to 0.6 b_i$) and at 80% compression ($b_i \to 0.2 b_i$). As a result of the compression, all allowed energy zones expand. The lower bound of the first two (lower) allowed zones noticeably shifts down the energy level, and their upper bound shifts upwards. The upper zones significantly expand, shifting even higher. At the same time, the forbidden zones are significantly narrowed.

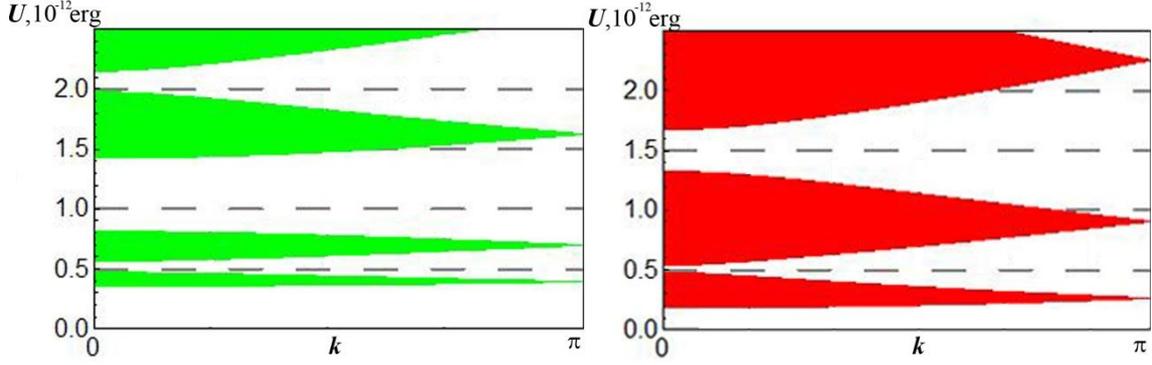

**Fig. 8.** The allowed energy bands at 40% compression and at 80% compression.

As each band expands during compression, the number of electrons that can occupy each such zone increases. As a result, part of electrons is shifted to the previous zone with lower energy levels. If initially it was, for example, a metal, then as a result of this transformation the electrons can completely release the previously partially filled band, and the band with lower levels of energy remains completely filled. As a result, we have a metal-dielectric transition. If initially it was a dielectric, then as a result of the expansion of the previously completely filled band, it will become only partially filled and the dielectric - metal transition will occur. There may be several such sequential transitions for some materials, that occur with increasing pressure. The number of transitions is finite (limited) and does not exceed the initial number of bands on which at least some number of electrons were present.

The pressure influences on fluorescence and on phosphorescence are of some interest. These effects can disappear under pressure, when the intermediate level (for fluorescence) or the metastable level (for phosphorescence) becomes the resonant energy level (when changing $L$). As the pressure further increases, these effects (fluorescence and phosphorescence) can appear and disappear. These effects can also appear and disappear for the waves with other frequencies.

There is another phenomenon, which is of some interest. This is the pressure influence on the magnetism of materials. The magnetism can disappear under high pressure, when the energy of localized ordered internal electrons, which provide the magnetism, becomes the resonant one (and electrons become collectivized). The nonmagnetic material, on the contrary, can become magnetic under pressure. These magnetic–nonmagnetic material transitions can occur consecutively.

As the first approximation, the three-dimensional crystal can be imagined as some atom structure, which is bounded by means of collective electrons in "resonant-percolative" directions. For any resonant direction there exists some resonant distance $L_j$ between barriers. In general, the energy of such a configuration will be minimal. Any attempt of extension, compression, shear or torsion of a crystal leads to changing at least one resonant distance, that is, the resonant level of electrons shifted from its usual position. Because of this change of electrons energy in a crystal, the elastic force appears. In principle, the crystal structure problem can be formulated as follows: it is necessary to find the resonant energies and distances. For the equilibrium structure, the total energy must be the minimum. Let us consider questions of the type of crystal bonds and of the crystal strength under pressure. Since bands are shifted under pressure, the type of crystal bonds can change successively. Since we cannot provide an arbitrary given distance $L$ between barriers, these transitions can occur with discontinuities.

Following the type of crystal bonds, the crystal can be successively changed (with $L$) from fusible to refractory one and *vice versa*. The oscillating changes of the anisotropy of some properties will take place in a linear compression (with $L$, but not with $P$). The distension cannot produce periodic changes in the properties of materials, since the rupture of a crystal will take place. From this viewpoint it is interesting to investigate the rupture of a metallic crystal (semiconductor, semimetal) with current (flowing across or along the crystal). In this case strengthening a crystal should occur, if the new resonant energy of crystal electrons under tension is higher than the old resonant energy. If the new resonant level is shifted to a lower position, the crystal strengthening decreases. From this point of view, it is evident that the explosion of a metallic crystal will take place under the action of a current with intensity larger than some critical one. With changes of the resonant energy of electrons, new regions with less strong bonds appear, that is, the separation into layers occurs. The investigation of transitions during compression of a crystal with current is of some interest (these transitions can take place under pressures other than in the usual case).

We are also interested in the following important fact. If two separate barriers are at such distance $L$, that $D_{12}=1$, then the total transmittance $D=1$ for any distance $b$ from this pair of barriers to the other similar pair of barriers. Therefore, the resonant tunneling can take place not only for periodic potentials, which are simple models of crystals, and we arrive at the conception of "barrier associations". All above considered consecutive (with $L$) transitions can occur for the classical crystal structure. For barrier associations we must consider $L$-changes within the association (changes in the distance between associations have comparatively small influence on material properties).

We begin with a crystal bond. According to our assumption, any crystal can be imagined as series of barriers, which are bound by two potentials: one binds the barriers together into associations (pairs, for example), the other one binds the associations of barriers into the crystal as a whole. If electrons, which hold the barriers to each other in the association, are located in one of potential wells most of the time, that is, we have the ionic or covalent bond (according to the degree of localization). The pure metallic bond can be accomplished by collective resonant electrons (there can be several of such bonds: for the associations and for the crystal as a whole). However, various types of bonds can exist for solids (and liquids) in reality. When this takes place, associations are always bound more firmly, than the crystal. The crystal growth occurs from a "micro level" (the origination of associations) to a "macro level" (the origination of a crystal as such).

From the above reasoning a number of experimental facts can be qualitatively understood, namely, such properties of materials as brittleness, ductility, strength, some properties of melts. As for the strength of a crystal, the rupture of a crystal under tension takes place between the barrier associations; that is why the strength of a real crystal differs from that of an ideal crystal.

Some material must possess the brittleness, provided that differences between the bonds in the association and between the associations in the crystal are negligible. The metallic materials often possess the ductility. In this case, bonds inside any association are considerably stronger than bonds between associations; therefore, forces acting on the crystal can shift associations without disruption. Usage of barrier associations is the other method of description only. (The usual method of description is the following: brittle − ductile transition is related to changes in the mobility and concentration of dislocations.) There is only mediated connection between the material strength and the energetic advantage of bonds: the latter depends on the total lowering of energetic levels of collective electrons, whereas a "weak point" in these bonds determines the strength.

There is an interesting example of metal, which is the liquid at standard conditions: it is mercury. For mercury, the thermal fluctuations are capable of destroying the crystal, whereas bonds in associations persist. As a result, the mercury remains to be a good conductor of electricity. Similarly, the melting of other metals implies the disruption of bonds between associations, whereas associations by themselves persist, and, because of this, despite the absolute disruption of a crystal lattice, the melt remains to be a good conductor of electricity; in this case, the electrical conductivity changes insignificantly. (Additionally, we note in the brackets that the BCS theory for superconductivity is designed for a perfectly ordered crystal. In the real superconductor, on the size of any Cooper pair, there is a large number of twin boundaries, dislocations, etc. So, resonant barrier associations could help here.)

In terms of this conception it becomes clear, why the transparent disordered media (glass) or the electrical conduction of disordered media (of alloys) are possible. If an association is transparent for the wave with a given energy, then a disordered material as a whole, which consists of such associations, will also be transparent (glass). If the energy of electrons is the resonant energy for an association, then the energy will also be resonant for a material as a whole (alloy).

The resistance of alloy is determined by thermal fluctuations and by mismatch in resonant energies of associations. Therefore, it is closely associated with a contact resistance (or contact potential). One can expect that at low temperature a little addition of one metal to the other one (dopant) causes just the same variations in resistance, as a little addition of second metal to the first one. That is, the residual resistances can be close to each other.

If the origination of a great number of contacts between various barrier associations is energetically advantageous, then we have an alloy. Otherwise, the alloy does not exist, and we have "the colloidal mixture". From this viewpoint, it is interesting to study the pressure influence on alloys. In this case, various periodic transitions can be observed, namely, "the metallic alloy − nonmetallic disordered material" transitions, the periodic fibration and the appearance of "a colloidal mixture". In addition, the transitions from an amorphous material to a crystal and even from the disordered material to a compound with unit stoichiometric formula are possible. It is also interesting in this case additionally to study the influence of a current on alloys: if the resonant levels approach each other, a stronger metallic bond arises. There can exist some critical current at which an alloy begins to separate into layers of various metals.

Factually, we assume above that all barriers are motionless. Since the transmittance of real barriers depends continuously on the wave vector (or on the energy of particles), some small variations in $k$ (or in $E$) produce only small deviations from $D = 1$. Since the light propagation is very fast process, temperature fluctuations will only slight change the solution $D = 1$. For mutual fluctuations of barriers or fluctuations of the association as a whole, this effect can be taken into account with using of the Debye−Waller factor.

Thus, the barrier associations can describe some disordered mediums (glasses, alloys, melts) without the notions of short-range and long-range order: we can have full correlation inside associations and no any correlation between associations.

## 7. Some remarks on 3-D models

For beginning and as some hypothesis, in two- and three-dimensional cases the movement of particles (or the propagation of waves) can take place on "the easiest way" principle. The movement in such a "resonant-percolative" trajectory is analogous to the one-dimensional case. For example, particle scattering for large speeds take place in the narrow cone (some analogy with one-dimensional case). Fractal structure also possesses some preferred directions (and sizes), and particle movements or property changes depend on the changes along the fractal chains.

Recall that in reality, one-dimensional case can include very thin linear objects (for example, thin wires), as well as those cases where there is a dependence of the solution on one coordinate only. Analogously, two-dimensional case can include thin films, as well as those cases where there is a dependence of the solution on two coordinate only. Such solutions can be sometimes implemented for real three-dimensional objects also.

The three-dimensional case of a diatomic molecule is some analog of the one-dimensional model of two barriers, discussed in Sect. 2. As an example, we can consider scattering of slow neutrons on hydrogen molecules (see [28, p. 591]). In this case, the neutron scattering cross section of the molecule has the form for orthohydrogen – orthohydrogen and parahydrogen – parahydrogen transitions:

$$d\sigma_n = \frac{4}{9}\frac{p'}{p}\left|\left(\cos\frac{\mathbf{qr}}{2}\right)_{0n}\right|^2 [(3f^+ + f^-)^2 + I(I+1)(f^+ - f^-)^2]do, \qquad (29)$$

and for orthohydrogen – parahydrogen (α=1) and parahydrogen – orthohydrogen (α=3) transitions:

$$d\sigma_n = \alpha\frac{4}{9}\frac{p'}{p}\left|\left(\sin\frac{\mathbf{qr}}{2}\right)_{0n}\right|^2 (f^+ - f^-)^2 do, \qquad (30)$$

where $\mathbf{p}$ and $\mathbf{p}'$ — are momentums of incident and scattered particles; $\mathbf{q} = (\mathbf{p}'-\mathbf{p})/\hbar$; $\pm\mathbf{r}/2$ — are the radius-vectors of two nucleuses in the molecule relative theirs center of inertia; matrix elements are calculated for the initial $\psi_0$ and final $\psi_n$ wave functions of the nuclear motion; $I$ is the full nuclear spin; $f^+, f^-$ — are the scattering amplitudes depending on the total spin of the nucleus and neutron; $do$ — is the solid angle element. Here we are interested in the fact that the dependencies in formulas (29) and (30) are periodic in $\mathbf{r}$, that is quite similar to the results obtained for the one-dimensional case from Sects. 2 and 6. So, the corollaries can be expected similar.

Even from the Bragg law for three-dimensional crystal

$$2d \sin \theta = n\lambda , \qquad (31)$$

we see that with fixed $\theta$ and $\lambda$, if we increase $d$ by a certain amount $\Delta d = \lambda / (2\sin\theta)$, then the choice instead of $n$ of the value $n + 1$ will again give the same identity. Similarly, reducing the value of $d$ by the value of $\Delta d$ together with replacing $n$ by $n - 1$ (until we will have $n = 1$) also gives the same result. Thus, we again have a successive change in the observed pattern (as for 1-D model). Similar successive changes occur in "the model of almost free electrons in a periodic potential field" for the 3-D case [29]. So, for the forbidden zone, we have the condition

$$k_x n_1 + k_y n_2 + k_z n_3 = \frac{\pi}{a}(n_1^2 + n_2^2 + n_3^2), \quad n_i \in Z, \quad i = 1,2,3, \qquad (32)$$

whence it is seen that the replacements $a = Na$, $n_i = n_i N$ (where $N$ — is natural number) lead to the same condition (32). For another limiting case, the "strong coupling approximation for electrons in a metal", we have a periodic dependence on $a$ (cubic lattice) for the allowed energy values:

$$E = E_0 - \alpha - 2\gamma(\cos k_x a + \cos k_y a + \cos k_z a), \quad \alpha, \gamma - \text{constant} . \qquad (33)$$

Thus, with respect to the properties of interest to us, the three-dimensional cases behave similarly to the one-dimensional cases considered by us. This allows us to hope that the hypotheses expressed may be applicable not only to model problems, but also to real materials. Although we understand that the quantitative application of qualitative hypotheses to real materials requires completely different methods and more rigorous calculations.

## 8. Conclusions

The exact solutions for some quantum-mechanical models are considered. This can be related to the enlightening of wave and quantum-mechanical barriers. The explicit equations for the transmission and reflection coefficients are derived. The use of exactly solvable models helps in science and education to better understand the theory under study. This article does not pretend to strictly substantiate all the hypotheses expressed. However, the explicit formulation of hypotheses allows us to organize a purposeful search for the expected phenomena. The considered description can relate not only to quantum mechanics and plasma physics, but also to the coated optics [30] and even to atmospheric physics. Wave equation can be applied here, and resonance concentration of waves can effect on the safety of flights. Collective processes take place in the atmosphere also, for example, atmospheric electricity, and can be explained with usage of quantum mechanics [31].

The quantum-mechanical model of two barriers can qualitatively describe some phenomena in solids. Such properties of materials as color, transparency for externally impinging particles or waves, as conductivity and other ones can oscillate (with $L$) under pressure. The same material can exhibit various properties: metallic and non-metallic, brittle and ductile, and so on. The transitions can occur consecutively under pressure.

Barriers in solids and liquids are combined into associations, where the bonds inside the association are stronger than those between associations. All material properties are determined by properties of an association. Changes between associations only insignificantly influence on material properties. From this viewpoint, some properties of disordered media, such as the transparency of glasses, the conductivity of alloys or melts, the brittleness and ductility, etc., can be understood. The 1-D model of three barriers is discussed. The 1-D model of three-well potential is applied to the phenomena under consideration. Some remarks on 3-D models are made.